\documentclass[twocolumn]{aastex631}

\shorttitle{PDRs4all: JWST NIRCam simulations}
\shortauthors{Canin et al.}

\usepackage[frozencache=true]{minted}

\begin{document}

\title{PDRs4all: Simulation and data reduction of JWST NIRCam imaging of an extended bright source, \\ the Orion Bar}

\author[0000-0002-7830-6363]{Am\'elie Canin}
\affiliation{Institut de Recherche en Astrophysique et Planétologie (IRAP), University of Toulouse, France. \texttt{olivier.berne@irap.omp.eu}}

\author[0000-0002-1686-8395]{Olivier Bern\'e}
\affiliation{Institut de Recherche en Astrophysique et Planétologie (IRAP), University of Toulouse, France. \texttt{olivier.berne@irap.omp.eu}}

\author{The PDRs4All ERS team}


\begin{abstract}

The James Webb Space Telescope (JWST) will be launched in December 2021, with four instruments to perform imaging and spectroscopy.
This paper presents work which is part of the Early Release Science (ERS) program ``PDRs4All'' aimed at observing the Orion Bar. It focuses on the Near Infrared Camera (NIRCam) imaging which will be performed as part of this project. The aim of this paper is to illustrate a methodology to simulate observations of an extended source that is similar to the Orion Bar with NIRCam, and to run the pipeline on these simulated observations. These simulations provide us with a clear idea of the observations that will be obtained, based on the ``Astronomer's proposal tool'' settings. The analysis also provides an assessment of the risks of saturation. The methodology presented in this document can be applied for JWST observing programs of extended objects containing bright point sources, e.g. for observations of nebulae or nearby galaxies. 

\end{abstract}



\section{Introduction} \label{sec:intro}

The James Webb Space Telescope (\citealt{gard2006}, JWST hereafter) is a space telescope to be launched on December 22, 2021. It was developed in an international collaboration of NASA, European Space Agency (ESA) and Canadian Space Agency (CSA), with four main scientific focuses: ``The End of the Dark Ages: First Light and Reionization"; ``The Assembly of Galaxies"; ``The Birth of Stars and Protoplanetary Systems"; and ``Planetary Systems and the Origins of Life". Thirteen Early Release Science (ERS) programs have been selected to demonstrate the scientific capabilities of JWST, and will provide public data to the community starting about 6 months after launch. In addition to the data, ERS projects are entitled to educate and inform the community regarding JWST's capabilities. The present papers is part of this effort in the context of the ERS program “PDRs4all: Radiative feedback from massive stars” (ID1288) which focuses on observations of the Orion Nebula \citep{ber21}. This 35-hour program will make use of three instruments aboard JWST, and will dedicate about 3 hours to imaging of the Orion Bar with NIRCam. The NIRCam instrument \citep{riek2005} has 29 filters between $0.5 \mu m$ and $5.0 \mu m$, out of which we will use 17 in the PDRs4all ERS project.  

In this paper, we present the methodology we have used to simulate NIRCam observations of the Orion Bar as planned in this ERS project. There are 3 goals for these simulations: 1) to obtain a clear and precise estimate of the field of view according to the settings we have selected in the Atronomer's Proposal Tool (APT\footnote{ \url{https://jwst-docs.stsci.edu/jwst-astronomers-proposal-tool-overview}}), 2) test potential saturation due to bright stars in the field of view, 3) obtain simulated data that we can run through the pipeline in preparation for the real data. For this purpose, we use the NIRCam simulator developed by the Space Telescope Sciences Institute (STScI) called Multi Instrument Ramp Generator (MIRAGE, \citealt{hil2019}), which allows us to simulate raw images. This paper is organised as follows: Section \ref{sec:instrument} gives an overview of the NIRCam instrument. In Section \ref{sec:mirage}, we present the MIRAGE simulator with examples of how it is used. Section \ref{sec:pipeline} presents the Data Reduction Pipeline. In Section \ref{sec:sat}, we discuss the impact of saturation on our images.

\section{NIRCam imaging of the Orion Bar} \label{sec:instrument}

\begin{figure}
    \centering
    \includegraphics[height = 8cm]{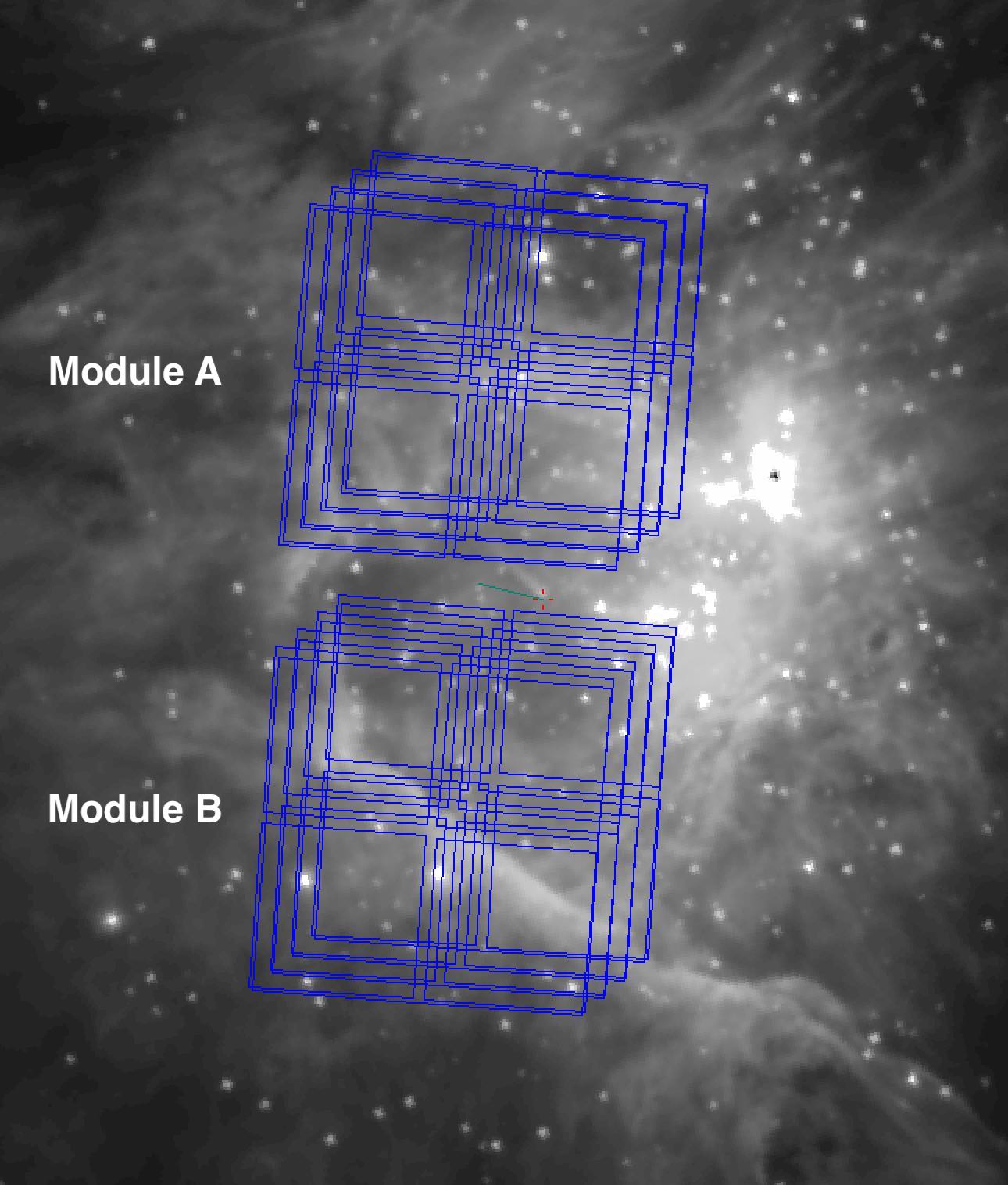}
    \caption{NIRCam field of view for the PDRs4all project as seen in APT, with Spitzer-IRAC \citep{faz04} Channel 1 (3.6$\mu$m) image in background. Modules A and B are indicated.}
    \label{fig:fig1}
\end{figure}

The Near Infrared Camera (NIRCam, \citealt{riek2005}) is one of the four JWST instruments. There are five observing modes with NIRCam. Here we are interested in the imaging mode\footnote{ \url{https://jwst-docs.stsci.edu/jwst-near-infrared-camera}}. \\
NIRCam has 10 different detectors separated in 2 modules (A and B, seen in Fig. \ref{fig:fig1}), assembled with gaps between them. There are 8 small detectors (A1, A2, A3, A4, B1, B2, B3 and B4, seen in Fig. \ref{fig:fig_hst}) for short wavelengths and two large detectors (A5 and B5) for long wavelengths. This allows us to simultaneously observe short and long wavelengths. NIRCam has 29 filters between $0.5 \mu m$ and $5.0 \mu m$.
As part of the PDRs4all ERS project, 7 NIRCam observations are planned combining long and short wavelength observations, for a total of 17 filters. Each exposure will have one integration and each integration will consist of 2 groups with 4 dithers giving a total integration time of $85.894s$. These settings can be viewed using the Astronomer's Proposal Tool (APT), by retrieving directly program 1288 from the file/retrieve from STScI menu.
The footprints of these NIRCam observations as seen with APT are shown in Figs~\ref{fig:fig1}-\ref{fig:fig_hst}.

\section{MIRAGE Simulation} \label{sec:mirage}

The JWST Multi Instrument Ramp Generator (MIRAGE), is a simulator to create JWST-like NIRCam images. It is an open source Python package developed by the STScI \citep{hil2019}. 
In the following sections, we describe the
steps to perform the simulation for the case of the Orion Bar observations as planned in the PDRs4all project. 

\subsection{Step 1: Preparing MIRAGE inputs}

MIRAGE works with several inputs: it requires APT generated files describing the observation, the roll angle of the observation, and, for the case of an extended source, a 2D image providing the spatial texture of this source, or {\it scene image}.


\paragraph{APT exported files} For the simulations, MIRAGE needs several files from APT. The xml and the pointing files are two files exported from the APT proposal. These files contain the observations, the filters used, the groups per integration, the dithers.

\paragraph{Roll angle} The roll angle of the telescope is needed so that MIRAGE can position the field of view correctly on the sky. It corresponds to the position angle of the V3 axis in degrees east of north. For a given date of observations, the roll angle can be calculated using the command in Listing \ref{lst:roll_angle}.  Here we have used the date of the planned ERS observations, i.e. Sept 10, 2022.

\begin{figure}
    \centering
    \includegraphics[height = 5cm]{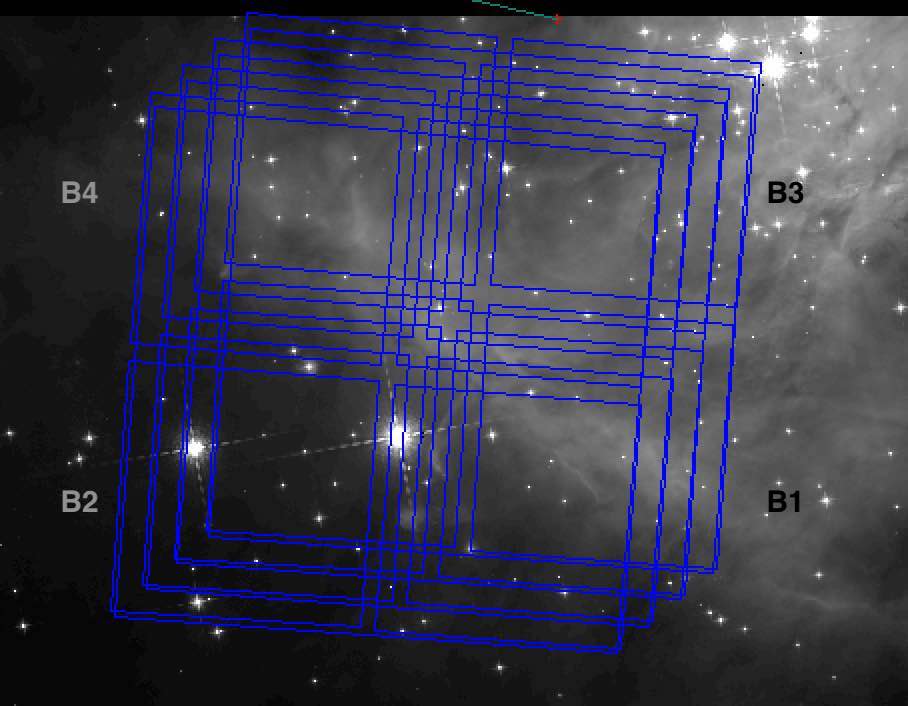}
    \caption{NIRCam field of view for the PDRs4all project as seen in APT, with HST-WFC3 \citep{kimble2008} with the F139M filter (1.4$\mu$m) image in background.}
    \label{fig:fig_hst}
\end{figure}

\paragraph{Scene image} We used two images for the input scenes.
At short wavelengths, we use the Hubble Space Telescope image of the Orion Nebula at 1.4 $\mu$m obtained by \citet{rob20} with the Wide Field Camera 3 (WFC3). This image is presented in Fig.~\ref{fig:fig_hst}. At longer wavelength, we use the 3.6 $\mu$m Spitzer-IRAC image (Fig. \ref{fig:fig1}) from \citet{meg15}. 
Information on astrometry in the headers of these images is used by MIRAGE, hence it is important that theses headers are written properly. 

\subsection{Step 2: Running the simulation}

The simulation has 4 stages: the creation of the YAML (YAML Ain't Markup Language) files, the generation of the seed image, the dark current preparation and the observation generation. We describe these steps hereafter. 

\paragraph{Creation of the YAML Files}

The first stage of the MIRAGE simulation is the creation of YAML files from the inputs above, especially the files exported from APT. YAML is a language used for configuration files as XML or JSON. The YAML files contain all the information needed for MIRAGE to work properly and simulate the right observations like the telescope and instrument settings, the reference files. 
The YAML generator allows to execute this first step. 
The YAML generator takes as inputs the xml and pointing files created with APT (see above), the roll angle, the datatype set as \texttt{raw} to obtain the same results as the JWST, and the output directories.
Listing \ref{lst:yaml} provides an example of how to create the YAML files.
The YAML files are generated in the \texttt{output\_dir} directory mentioned by users on local machine. There is one YAML file created per dither and detector. 
There are 8 detectors for short filters (Ai, Bi for i in \{1,2,3,4\}) and 2 detectors for long filters (A5, B5)
on NIRCam, and since in the PDRs4all project, there are 7 observations, each one with 4 dithers, this yields 
a total of $7\times4\times10 = 280$ YAML files for this project.

\begin{figure}
    \centering
    \includegraphics[height = 6cm, angle =270]{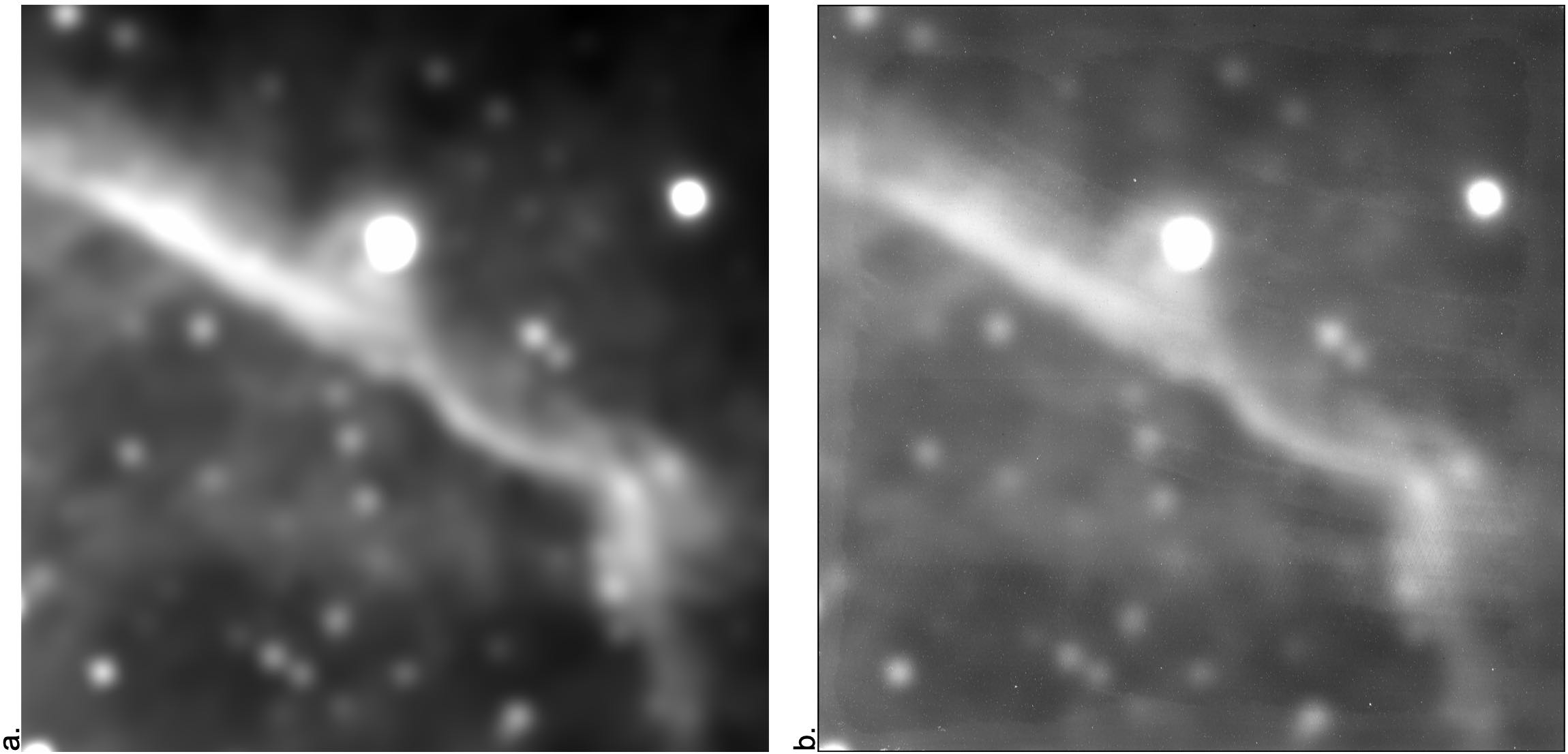}
    \caption{a. Seed image (\texttt{\_blotted\_seed\_image}) created by MIRAGE for detector B5 using the Spitzer image of the Orion Bar. b. MIRAGE simulated raw image (\texttt{\_uncal}) for the B5 detector with filter F335M.}
    \label{fig:mirage_f335m}
\end{figure}

\begin{figure*}
    \centering
    \includegraphics[height = 8.5 cm]{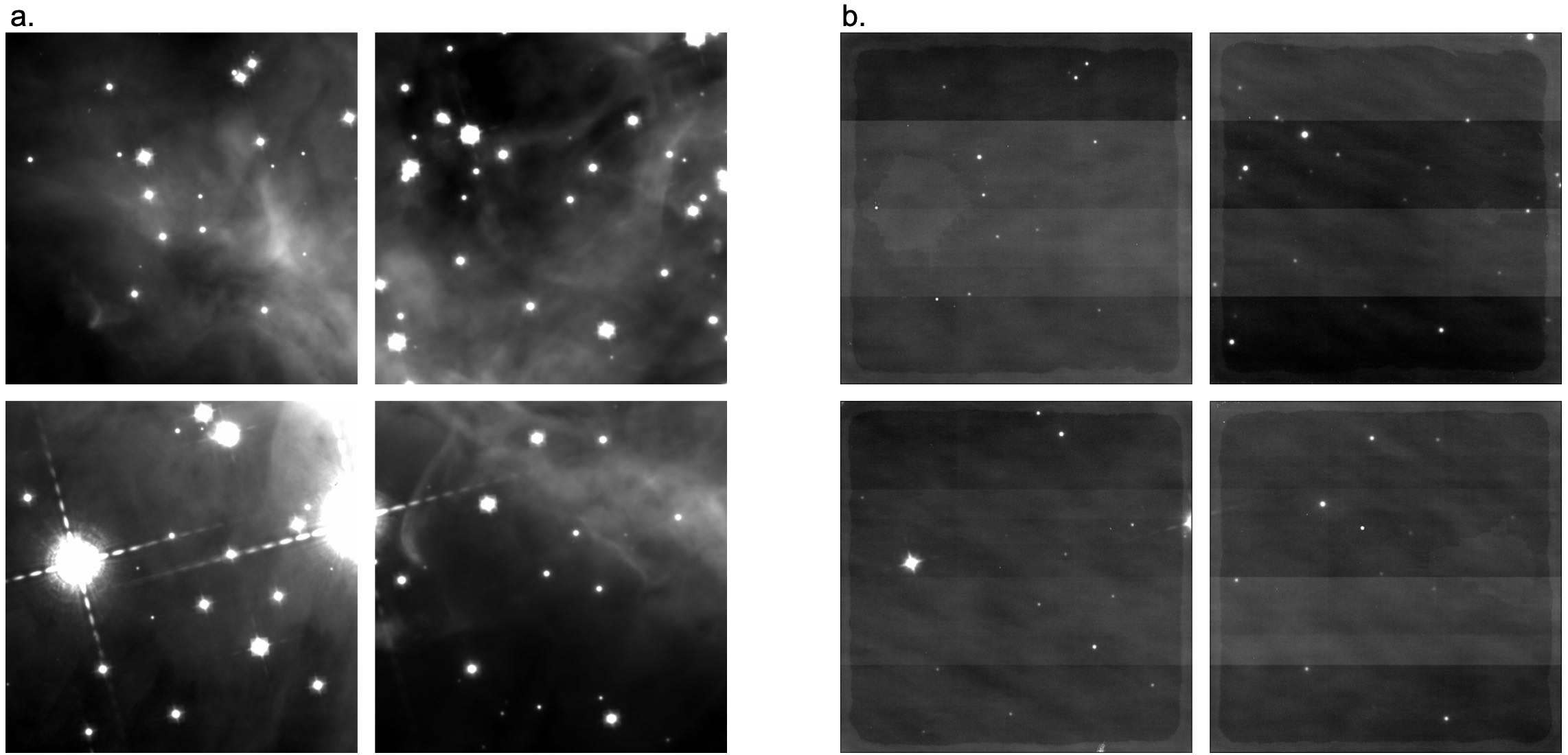}
    \caption{a. Seed image (\texttt{\_blotted\_seed\_image}) created by MIRAGE for detectors B1, B2, B3, B4 using the HST image of the Orion Bar. b. MIRAGE simulated raw image (\texttt{\_uncal}) for B1, B2, B3, B4 detectors with filter F140M.}
    \label{fig:fig5}
\end{figure*}

\paragraph{Creation of a seed image} The seed image corresponds to an idealized, noiseless image of the scene. As our extended images are bigger than the detectors, MIRAGE crops the scene. This stage produces an image with the extension \texttt{\_blotted\_seed\_image}. Listing \ref{lst:seed_image} shows how to create the seed image from the scene image. Figs. \ref{fig:mirage_f335m}a and \ref{fig:fig5}a show the results of this first step when applied to the scene images of the Orion Bar selected here for filters F335M and F140M, respectively.

\paragraph{Dark current preparation} This stage prepares the dark current exposure for the observation. This stage produces an image with the extension \texttt{\_dark\_prep\_object}. Listing \ref{lst:dark} presents the code to execute this task.

\paragraph{Observation generation} This stage produces the final raw image with the noise from the background but also due to the detectors. This stage combines the seed image and the dark current exposure and produces an image with the extension \texttt{\_uncal}. Listing \ref{lst:observation} provides an example on how to create the final observation from MIRAGE.

The MIRAGE simulation can take only one YAML file at a time, hence to run a full simulation on the 280 YAML files requires to write a for loop which includes the 3 previous stages, seed image, dark current preparation and observation generation.

Figs. \ref{fig:mirage_f335m}b. and \ref{fig:fig5}b show the resulting \texttt{\_uncal} images of the Orion Bar obtained for filters F335M and F140M, respectively. 
The readout pattern noise of NIRCam is clearly visible in these images as stripes, as is the additive photon noise.




\section{Reduction of the MIRAGE simulated images with the JWST Pipeline} \label{sec:pipeline}

\begin{figure*}
    \centering
    \includegraphics[height = 4 cm]{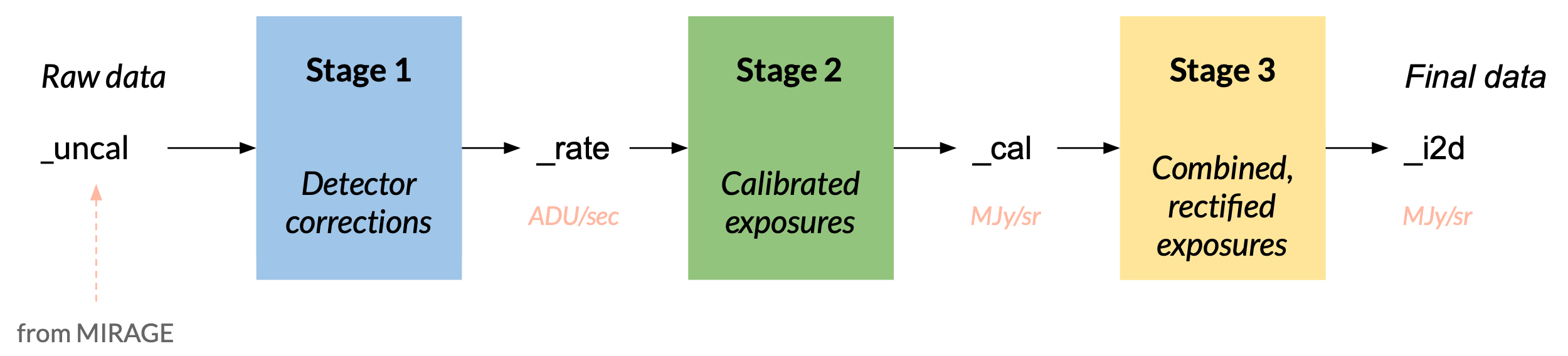}
    \caption{Overview of the steps of the JWST pipeline.}
    \label{fig:fig_pipeline}
\end{figure*}

The JWST Data Reduction Pipeline (``Pipeline" hereafter) is a Python package developed by the STScI. This software allows us to produce calibrated and reduced data from raw data taken by the JWST. From the raw data to the final data, the pipeline is composed of three stages (Fig. \ref{fig:fig_pipeline}).
In the following sections, we describe the three stages and code for processing the uncalibrated 
data to obtain Stage 3 data,
applied to the raw images created in the previous section using MIRAGE. The step-by-step description provided hereafter concerns the images obtained with detector B4 and filter F140M, however we also describe the results obtained applying the same methodology for the F335M filter.  

\subsection{Stage 1: Detector corrections}

\begin{figure}
       \centering
        \includegraphics[height = 17 cm]{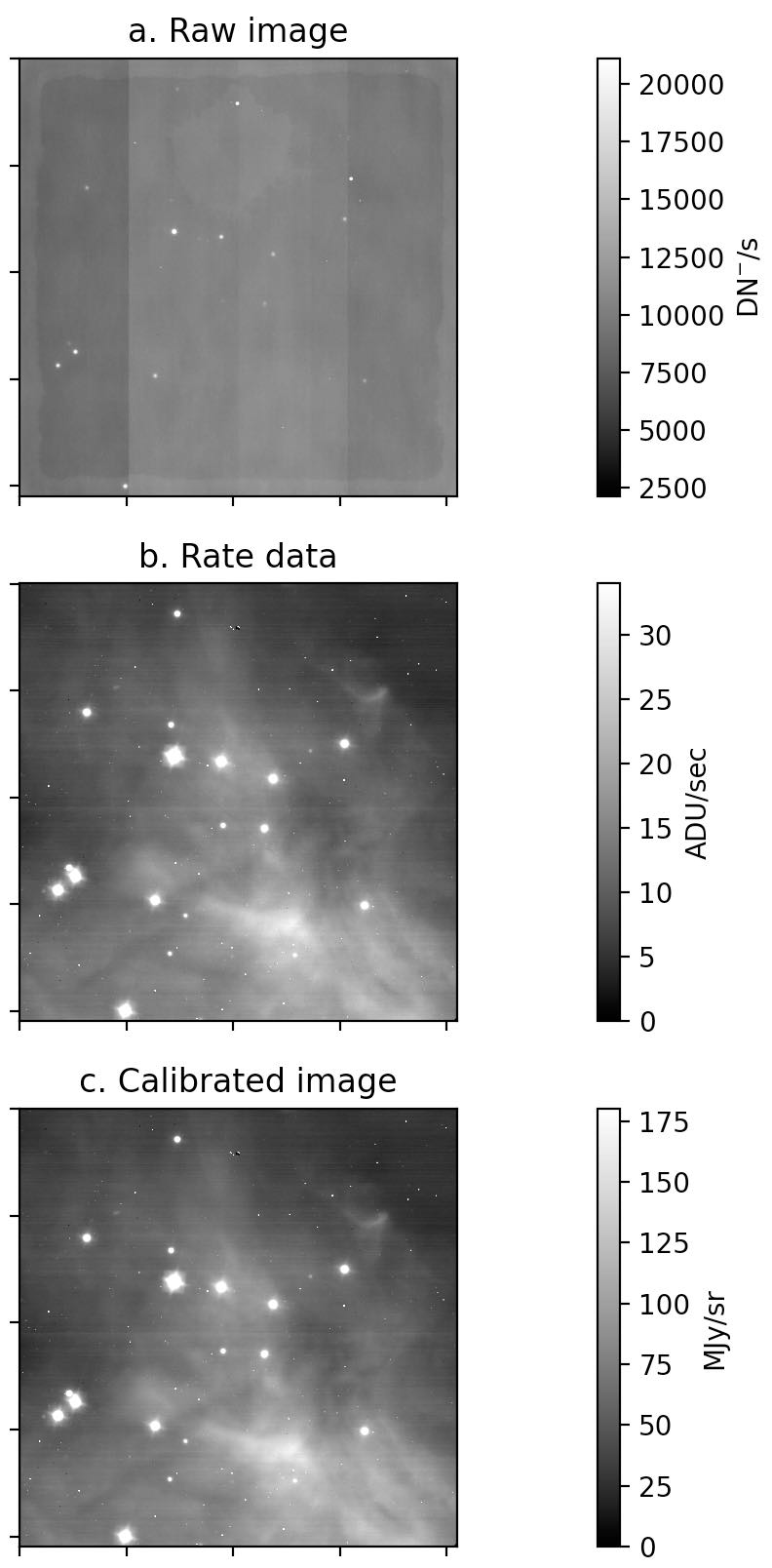}
        \caption{Processing of the simulated NIRCam image for the F140M filter, for one dither on one detector, from stage 1 to stage 2 with the JWST pipeline. a. Input raw image (\texttt{\_uncal}), b. rate data (\texttt{\_rate}), c. output calibrated image (\texttt{\_cal}).}
        \label{fig:fig9}
\end{figure}

This stage is the first one for all of the instruments. It corrects a part of instrument signatures in particular by removing the readout pattern (the stripes). It is called \texttt{CALWEBB\_DETECTOR1} or \texttt{Detector1Pipeline} in the Python pipeline. The input of this stage is a \texttt{RampModel} or an \texttt{\_uncal} file corresponding to a single raw exposure. The output of MIRAGE can be used here if during the simulation, datatype is set to \texttt{raw}. This stage only takes one file at a time and returns uncalibrated slope images in units of ADU/sec. It produces a \texttt{\_rateints} file, which is a 3D product with the results of each integration and a \texttt{\_rate} file which is a 2D product corresponding to the average of all the integrations in an exposure. Listing \ref{lst:stg1} presents the code to perform this stage. In Fig. \ref{fig:fig9} (a-b) we show the effect of applying this step on the raw F140M image created with MIRAGE. It can be seen that the stripes from the readout pattern have been removed, but the image is still not calibrated and there are residuals of the instrumental signature.

\subsection{Stage 2: Calibration of exposures}

This stage corrects other instrumental signatures and calibrates the exposures. This step is performed using the \texttt{CALWEBB\_IMAGE2} or \texttt{Image2Pipeline} in the Pipeline. The input of this stage is an \texttt{ImageModel} or a \texttt{\_rate} file, corresponding to the output of Stage 1. This stage returns calibrated (\texttt{\_cal}) but unrectified slope images in units of MJy/sr. The corresponding files have the extension \texttt{\_cal}. The code to perform this stage is presented in Listing \ref{lst:stg2}. Fig. \ref{fig:fig9} (b-c) presents the calibration of the image. 

\subsection{Stage 3: Combined and rectified exposures}

This stage combines multiple exposures from dithers or mosaics and rectifies the exposures to produce one unique mosaic, the final product. In this stage, we use the imaging mode called \texttt{CALWEBB\_IMAGE3} or \texttt{Image3Pipeline} in the Pipeline. The input of this stage is an \texttt{ImageModel} or a \texttt{\_cal} file corresponding to the output of Stage 2. To combine multiple exposures, an association (ASN) file is created which contains the images to be combined. Any combination of detectors and dithers is possible. To be read by the pipeline, the ASN file must be transformed into a \texttt{ModelContainer}. The creation of the ASN file and how to read into a \texttt{ModelContainer} is presented in Listing \ref{lst:stg3_part1}, while Listing \ref{lst:stg3} presents the final steps of stage 3 performed after.
This returns the final mosaic image, rectified, 
in units of MJy/sr with extension \texttt{\_i2d}.  We note that in stage 3, contrary to stage 1 and 2, we turn off the \texttt{tweakreg} option
(Listing \ref{lst:stg3}). With the \texttt{tweakreg} option turned off, the first two images are aligned, then the third one is aligned with the previous combination, etc. This allows for a better alignment between combined images. Alternatively, if the \texttt{tweakreg} option is on, during the alignment, all images are aligned with the first one, but since there is  very little overlap between the first and the last dither, alignment is poor.

\begin{figure*}
    \centering
    \includegraphics[height = 6cm]{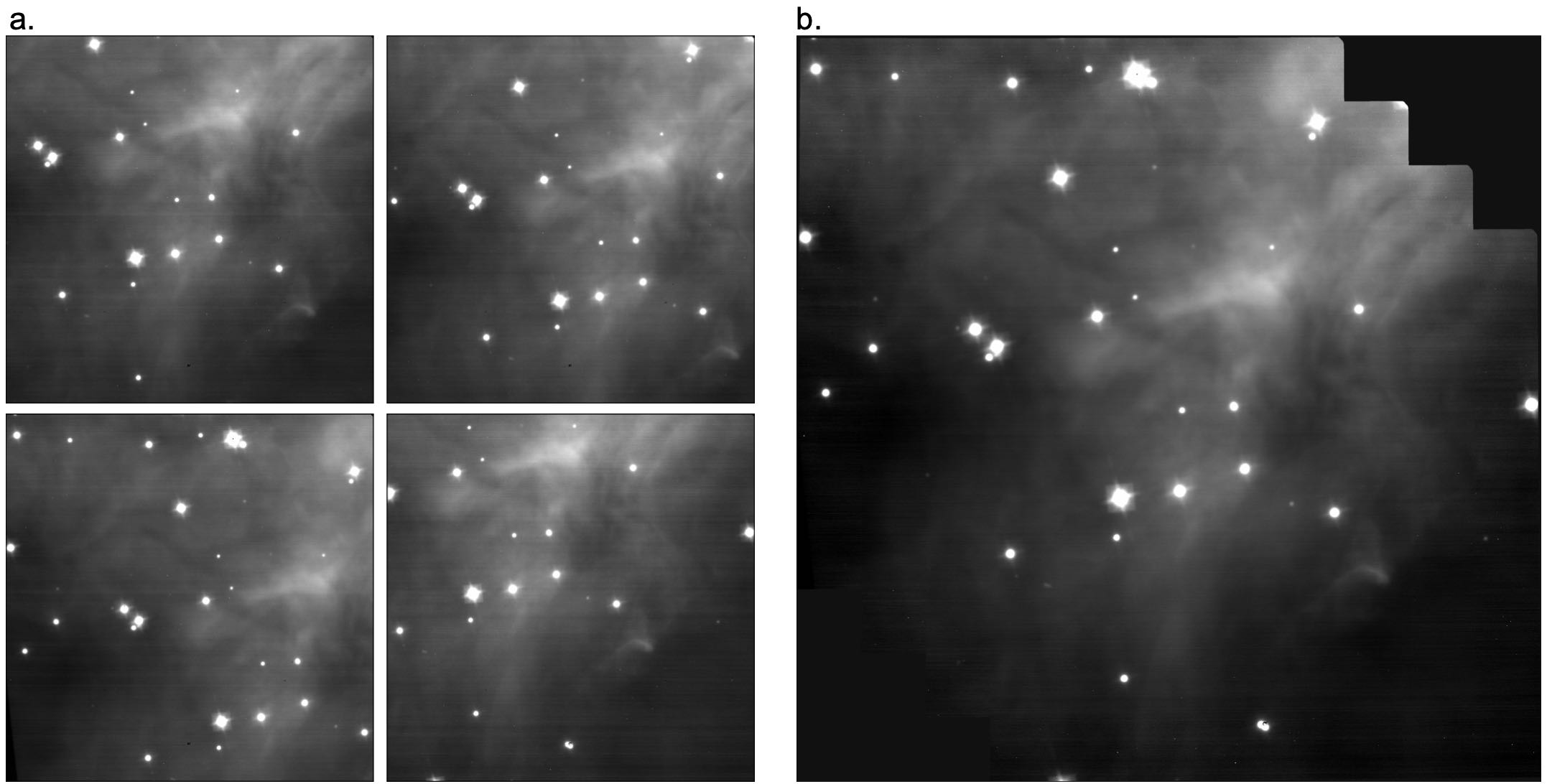}
    \caption{a. The 4 calibrated (\texttt{\_cal}) dithers for the detector B4 with the filter F140M. b. The 4 dithers combined in the final mosaic (\texttt{\_i2d}) for the detector B4 with the filter F140M.}
    \label{fig:stage3}
\end{figure*}

Fig. \ref{fig:stage3} presents the application of this step to the calibrated images of the 4 dithers of the same observation for detector B4, which are rectified and combined to produce the final image of the detector B4. 

\subsection{Final results} \label{subsec:results} 

\paragraph{Filter F140M}

\begin{figure*}
    \centering
    \includegraphics[height = 5cm]{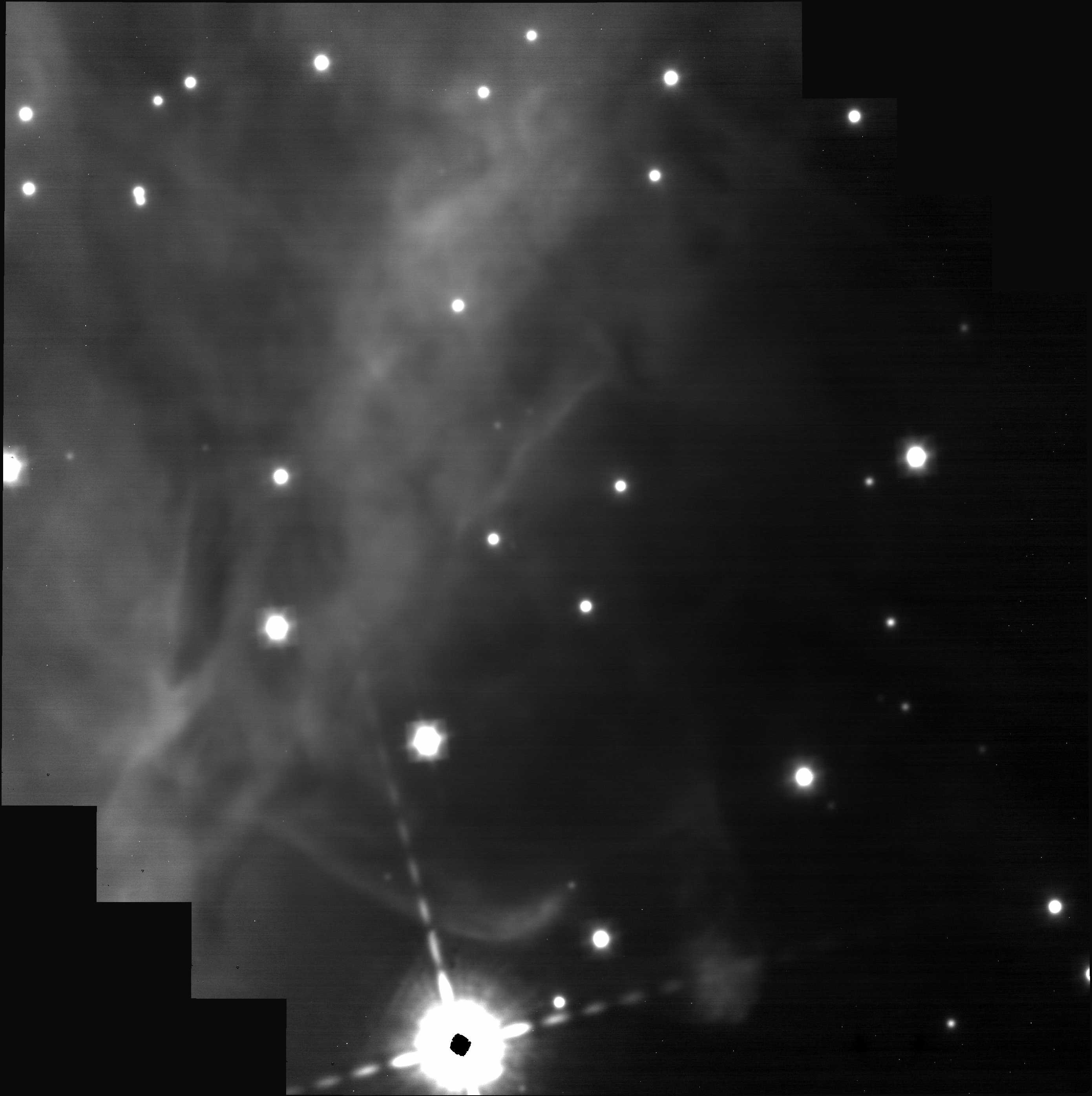}
    \includegraphics[height = 5cm]{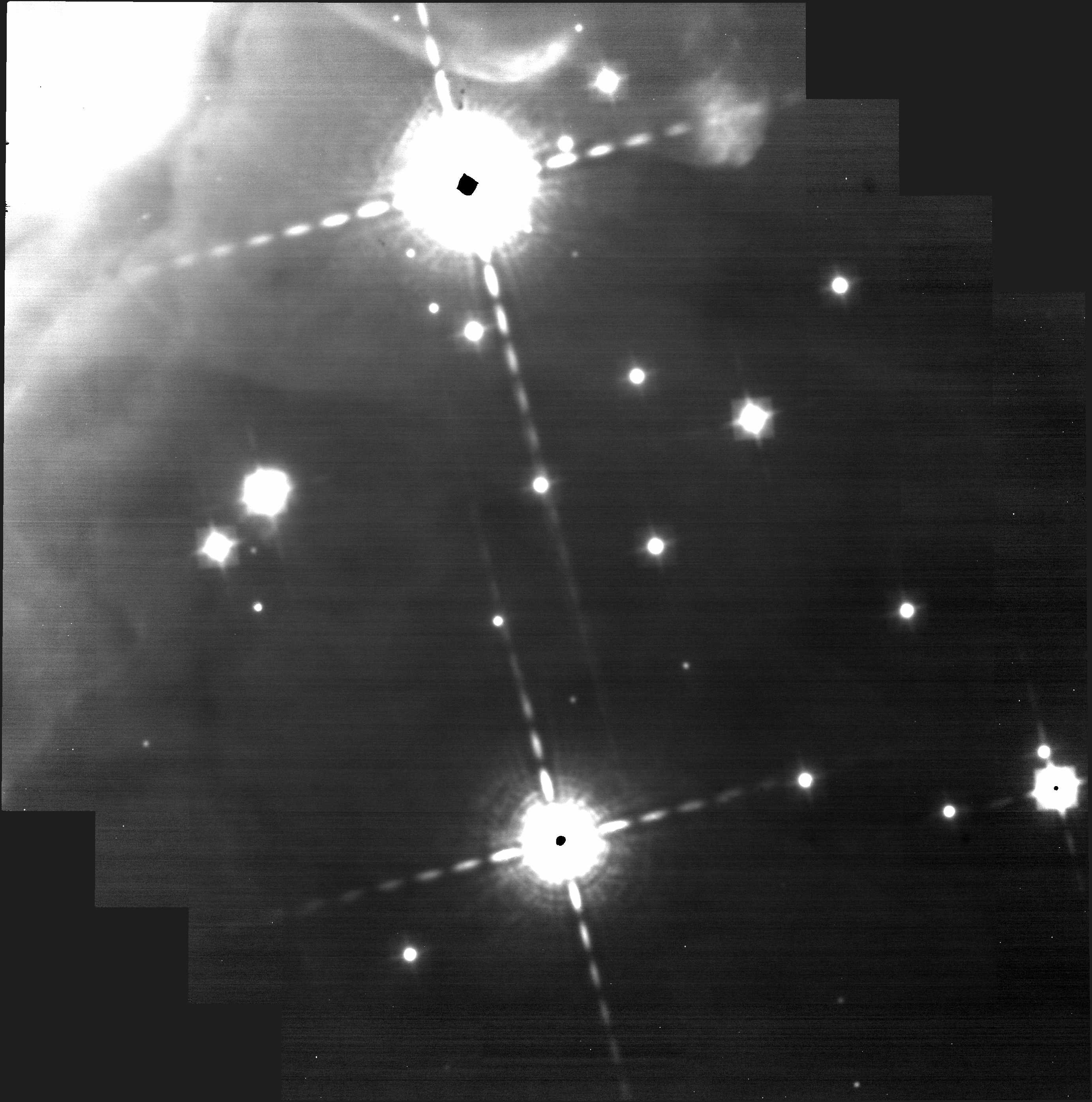}
    \break
    \includegraphics[height = 5cm]{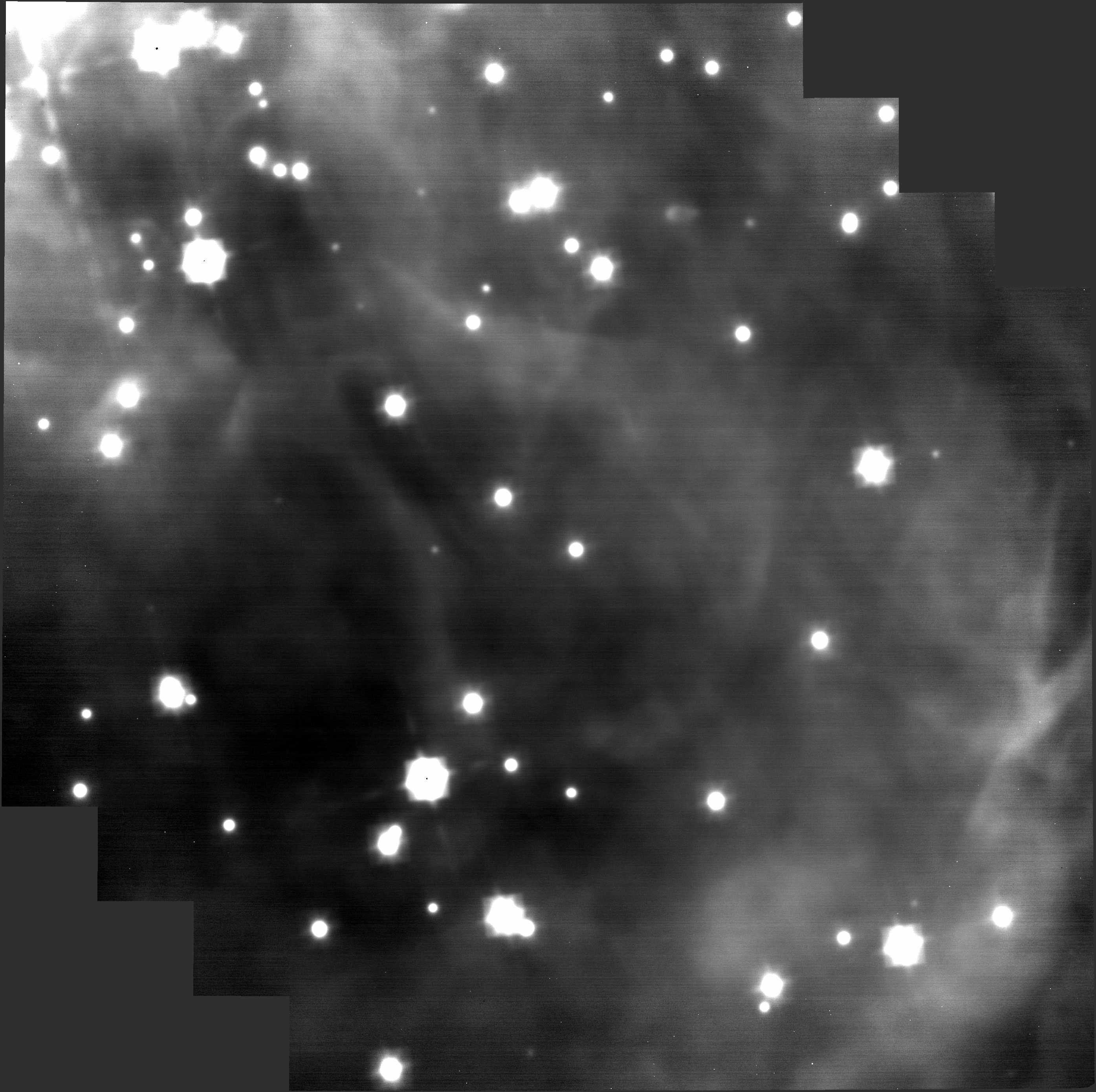}
    \includegraphics[height = 5cm]{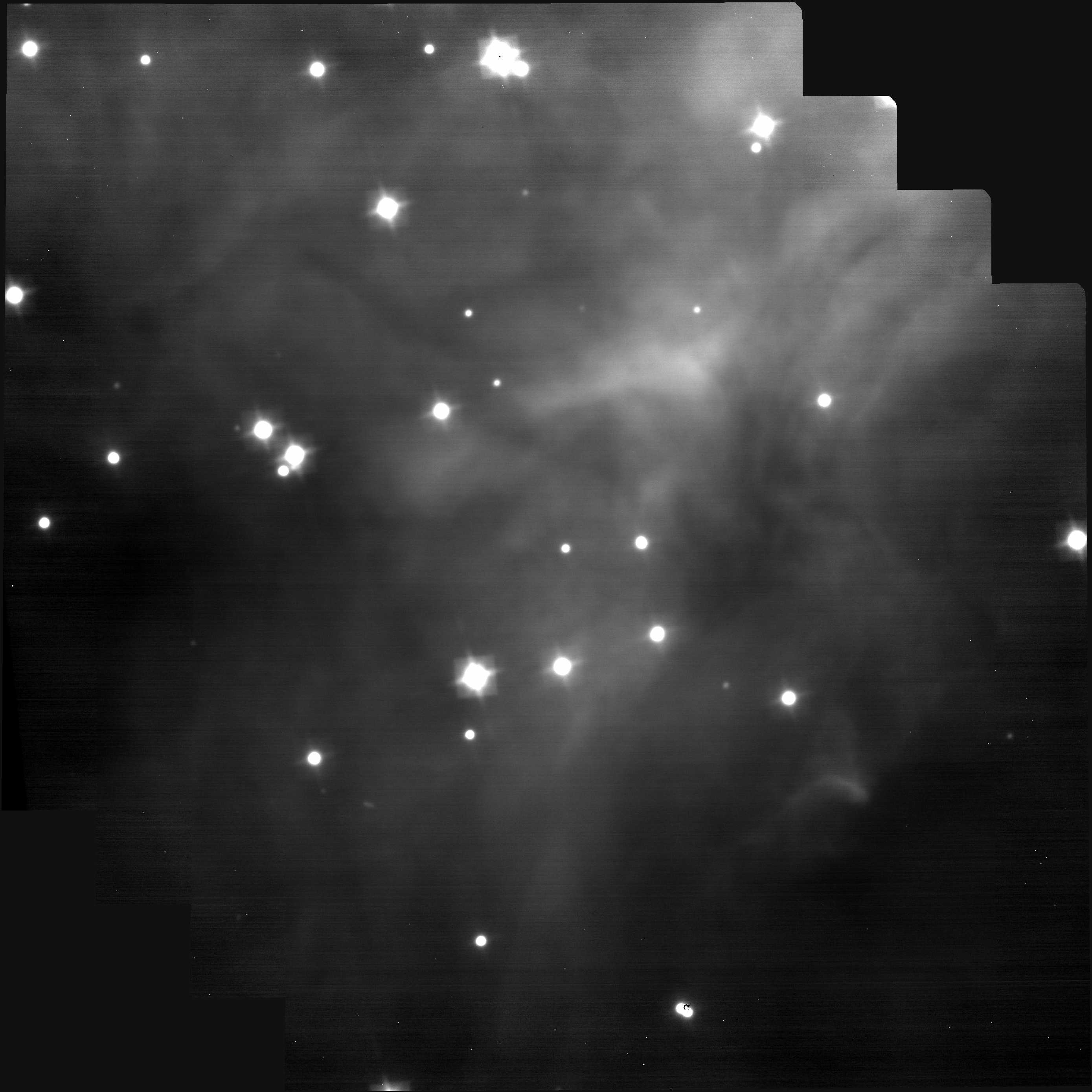}
    \caption{Dithers combined (\texttt{\_i2d}) for each detector with the filter F140M. From left to right and from top to bottom: B1, B2, B3, B4.}
    \label{fig:fig12}
\end{figure*}

\begin{figure*}
    \centering
    \includegraphics[height = 9cm]{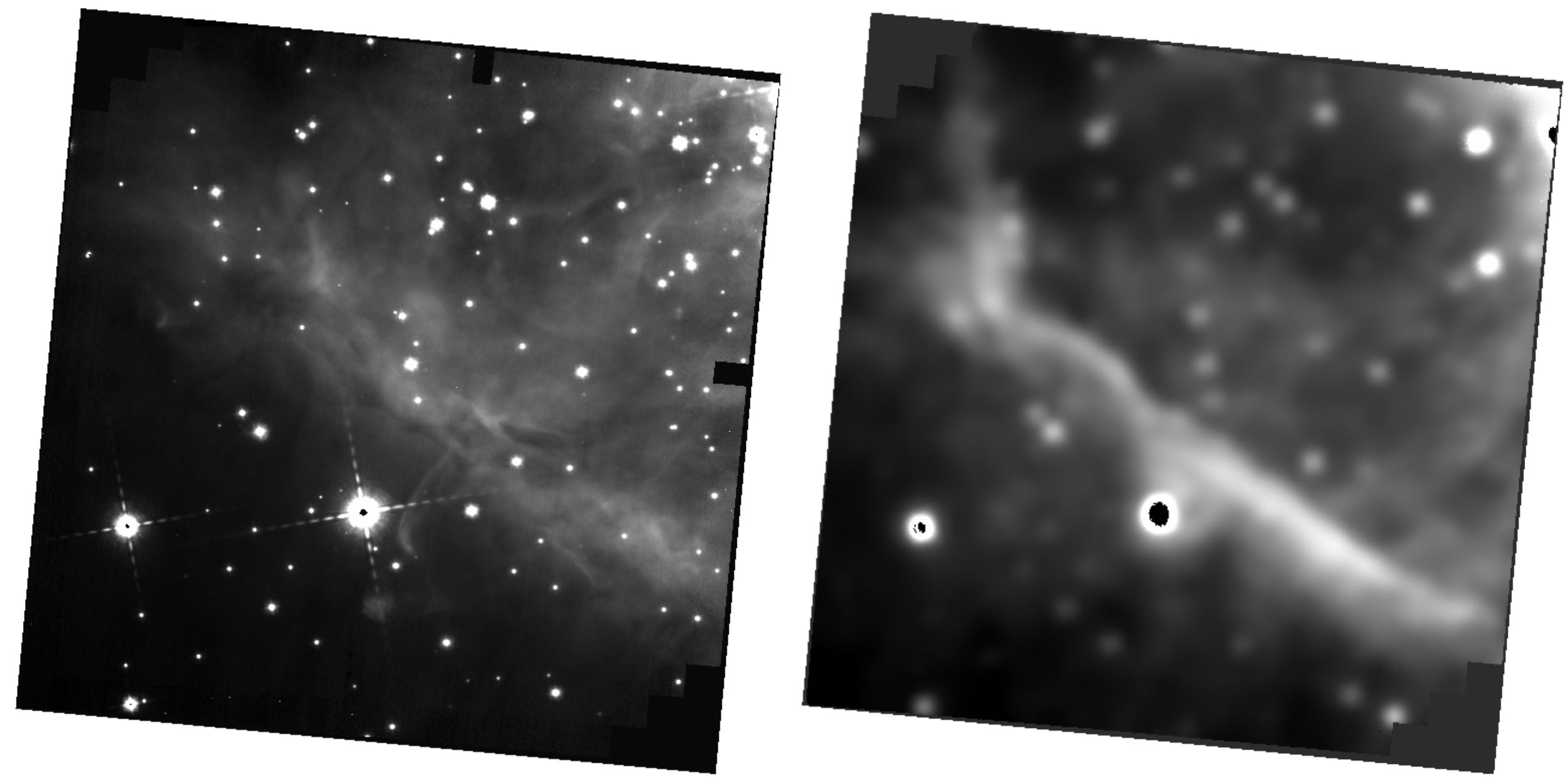}
    \caption{Final combined mosaic (\texttt{\_i2d}) using MIRAGE simulation and the JWST Pipeline. Left with the filter F140M for all dithers and detectors (B1, B2, B3, B4) and right  with the filter F335M for all dithers of the detector B5.}
    \label{fig:final}
\end{figure*}

As for Fig. \ref{fig:stage3}, the dither pattern can be seen on Fig. \ref{fig:fig12} which presents the calibrated and rectified mosaic for each detector in short wavelengths.  The final mosaic (Fig. \ref{fig:final} left) is obtained by combining all the images from the pipeline for each dither and detector, themselves obtained from the MIRAGE simulation. On this final image, saturation of the brightest stars can be seen and corresponds to the handful of black pixels at the centers of the brightest stars. 



\paragraph{Filter F335M}



Fig. \ref{fig:final} (right) presents the final mosaic for the F335M filter. As this filter is one for long wavelengths, 
the image is obtained using only one detector, B5. Hence the mosaic image only combines 4 dithers. This final mosaic
appears blurred. This is because of the low resolution ($\sim$ 2'') of the scene image that is used, as compared to 
the resolution of NIRCam ($\sim$ 0.2''). Saturation (black pixels) appear on the bright stars.





\begin{figure*}
    \centering
    \includegraphics[height = 5cm]{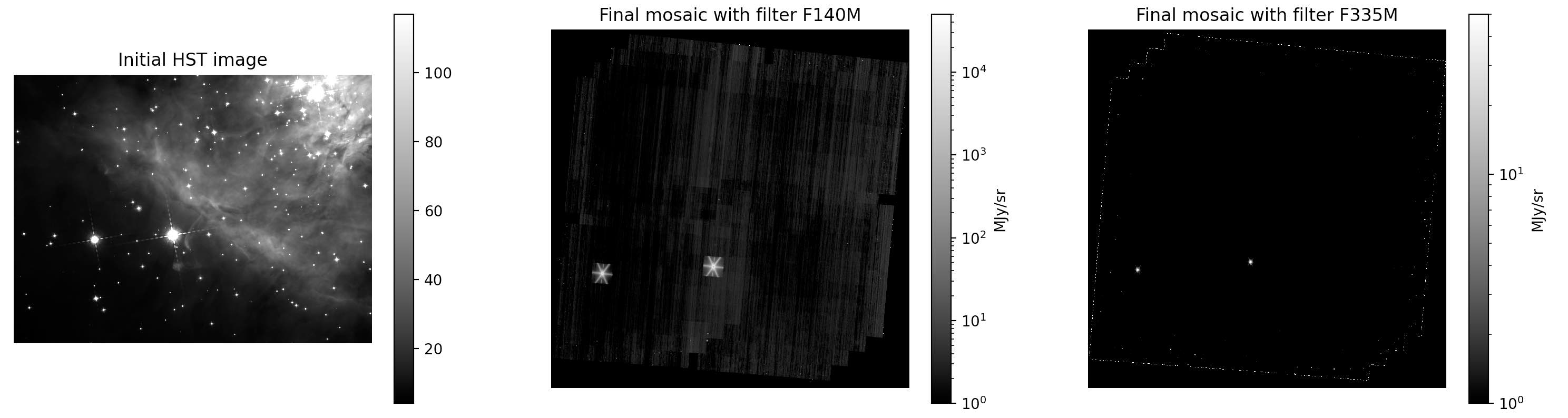}
    \caption{From left to right: initial HST image; final mosaic (\texttt{\_i2d}) of the MIRAGE image generated with the point sources catalog for the filter F140M; final mosaic (\texttt{\_i2d}) of the MIRAGE image generated with the point sources catalog for the filter F335M.}
    \label{fig:fig20}
\end{figure*}


\section{Evaluation of saturation and diffraction}\label{sec:sat}

One of the objectives of the simulations is to test the risks of saturation and/or contamination of the images due to diffraction 
patterns as a result of the presence of bright point sources in the field of view. As seen in the previous section, bright stars that are
present in the Orion Bar appear to saturate the NIRCam detector and produce diffraction patterns. However the effects of
saturation and diffraction contamination cannot be fully examined using the scene images introduced hereabove, because these 
have a resolution that is similar or smaller than the resolution of JWST. Hence, point sources present in theses scene images 
have been blurred by a point spread function (PSF) that is larger than that of JWST. The PSF are 0.048 arcsec for NIRCam at $1.4 \mu$m and 0.111 arcsec for NIRCam at $3.35 \mu$m. The HST WFC3 PSF is 0.143 arcsec at $1.4 \mu$m. The Spitzer IRAC PSF is between 1.7 and 2 arcsec.
We therefore perform an additional analysis, which consists of creating a seed image using a simple catalog containing 
the brightest two stars in the field of view, $\theta^{2}$ Ori A and $\theta^{2}$ Ori B. The catalog file contains the coordinates
of the stars and their magnitudes (retrieved from the Simbad database). $\theta^{2}$ Ori A is located at R.A. = 5:35:22.90225, decl. = -5:24:57.8172 and $\theta^{2}$ Ori B at R.A. = 5:35:26.40042, decl. = -5:25:00.7925. The magnitudes at 1.4$\mu$m and $3.35\mu$m for $\theta^{2}$ Ori A are 5.03 and 11.7, and for $\theta^{2}$ Ori B are 6.285 and 11.7. We re-perform the full simulation and data-reduction 
using this catalog as input, for filters F140M and F335M. Fig. \ref{fig:fig20} presents the results of this simulation. 
It can be seen in these images that saturation will only be localized on a few pixels, and that diffraction patterns 
will only affect a small region of the image. These effects are not critical for the PDRs4all objectives.

\section{Conclusion}

We have presented a methodology to perform simulations of observations of extended sources with NIRCam. 
This methodology relies on the use of previous images of the extended source to be observed with JWST,
combined with the use of the MIRAGE software, and the JWST Pipeline. In addition, we provided an example 
to test the effect of the presence of bright point sources in the field of view on saturation and diffraction 
patterns that can affect the image. This methodology can be applied by observers who wish to assess the quality of their 
observations of extended sources with NIRCam before they are executed, and can thus help optimize the 
planning of these observations.


\bibliography{sample631}{}
\bibliographystyle{aasjournal}

\appendix

\section{Listings}

\subsection{Command to calculate the roll angle in MIRAGE using the pointing file created with APT.}\label{lst:roll_angle}
\begin{minted}[frame=single]{python} 
yaml_generator.all_obs_v3pa_on_date(pointing_file, date=date) 
\end{minted} 

\subsection{Code to create the YAML files in MIRAGE}\label{lst:yaml}
\begin{minted}[frame=single]{python} 
yam = yaml_generator.SimInput(xml_file, pointing_file, verbose=True,
                              output_dir=output_dir,
                              simdata_output_dir=simdata_dir,
                              roll_angle=roll_angle,
                              datatype='raw'
                              )
yam.use_linearized_darks = True
yam.create_inputs()
\end{minted}

\subsection{Creation of the seed image from the scene image (\texttt{mosaicfile}).}\label{lst:seed_image}
\begin{minted}[frame=single]{python} 
with open(yfile) as file_obj:
    params = yaml.safe_load(file_obj)
# Define output filenames and directories
sim_data_dir = params['Output']['directory']
simulated_filename = params['Output']['file']
crop_file = simulated_filename.replace('.fits', '_cropped_from_mosaic.fits')
crop_file = os.path.join(sim_data_dir, crop_file)
blot_file = simulated_filename.replace('.fits', '_blotted_seed_image.fits')

# Crop from the mosaic and then resample the image
seed = ImgSeed(paramfile=yfile, mosaic_file=mosaicfile, cropped_file=crop_file,
               outdir=sim_data_dir, blotted_file=blot_file, mosaic_fwhm=0.009,
               mosaic_fwhm_units='arcsec', gaussian_psf=True)
seed.crop_and_blot()
\end{minted}

\subsection{Creation of the dark current}\label{lst:dark}
\begin{minted}[frame=single]{python} 
# Run dark_prep
dark = DarkPrep()
dark.paramfile = yfile
dark.prepare()
\end{minted}

\subsection{Creation of the final observation}\label{lst:observation}
\begin{minted}[frame=single]{python} 
# Run the observation generator
obs = Observation()
obs.paramfile = yfile    
obs.seed = seed.seed_image
obs.segmap = seed.seed_segmap
obs.seedheader = seed.seedinfo
obs.linDark = dark.prepDark
obs.create()
\end{minted}

\subsection{Stage 1 of the pipeline}\label{lst:stg1}
\begin{minted}[frame=single]{python} 
for f in uncal_files:
    Detector1Pipeline.call(f, save_results=True, output_dir=output_dir)
    
#uncal_files is a list that contains the paths of the raw files.
#save_result is set at True to save the _rate file in output_dir.
\end{minted}

\subsection{Stage 2 of the pipeline}\label{lst:stg2}
\begin{minted}[frame=single]{python} 
for f in rate_files:
    Image2Pipeline.call(f, save_results=True, output_dir=output_dir)
    
#rate_files is a list that contains the paths of the rate files.
#save_result is set at True to save the _cal file in output_dir.
\end{minted}

\subsection{Creation of the ASN file: first part of the Stage 3 of the pipeline}\label{lst:stg3_part1}
\begin{minted}[frame=single]{python}
# Creation of ASN file named IMA_asn.json
call(["asn_from_list", "-o", "./IMA_asn.json"]+cal_files+["--product-name", "ima_dither"])
# Transformation of the ASN file in ModelContainer
dm_3_container = datamodels.ModelContainer("IMA_asn.json")

#cal_files is a list that contains the paths of the calibrated files that will be combined.
\end{minted}

\subsection{Stage 3 of the pipeline}\label{lst:stg3}
\begin{minted}[frame=single]{python} 
nircam3 = Image3Pipeline()
nircam3.output_dir = output_dir
nircam3.save_results = True
nircam3.tweakreg.skip = True

stage3_output = nircam3.run(dm_3_container)

#save_result is set at True to save the _i2d file in output_dir.
\end{minted}



\end{document}